\documentclass[conference]{IEEEtran}
\IEEEoverridecommandlockouts

\usepackage{amsmath}
\usepackage{stmaryrd} 
\usepackage[ruled,vlined,linesnumbered]{algorithm2e}
\usepackage{subcaption}
\usepackage{hyperref}
\hypersetup{hidelinks=true, }
\usepackage[nameinlink]{cleveref}
\usepackage{fontawesome5}
\usepackage{graphicx}
\usepackage{amssymb}
\usepackage{etoolbox}
\usepackage[para]{footmisc}
\usepackage[table,xcdraw]{xcolor}

 \newbool{showcomments}
 \setbool{showcomments}{true}
 \ifthenelse{\boolean{showcomments}}
 { \newcommand{\mynote}[3]{
    \fbox{\bfseries\sffamily\scriptsize#1}{\small$\blacktriangleright$\textsf{\emph{\color{#3}{#2}}}$\blacktriangleleft$}
    }}
 { \newcommand{\mynote}[3]{}}

\newcommand{\awaitingreview}[1]{\faIcon{balance-scale}~}

\newcommand{\wontdo}[1]{\faIcon{times}~}
\newcommand{\sysname}{Collective Incentives\xspace}
\newcommand{\figpathsysmodel}{"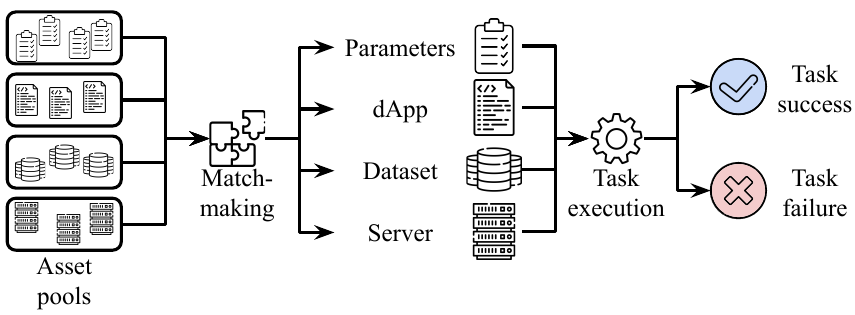"}
\newcommand{\figpathattack}{imports/img/DCMI-AttackModel.pdf}
\newcommand{\figpathruin}{imports/img/ruin_probabilities-2025-01-18-01-45-51.pdf}

\newcommand{\figpathindivfailrate}{imports/img/individual_failure_rate-Coll-SR-10000-100-0.05-10-0.1-20250118-110832.pdf}
\newcommand{\figpatheffecfailrate}{imports/img/effective_failure_rate-Coll-SR-10000-100-0.05-10-0.1-20250118-110832.pdf}
\newcommand{\figpathsysfailrate}{imports/img/system_failure_rate-ALL-100-0.01-10-0.05-20250117-234642.pdf}

\def\BibTeX{{\rm B\kern-.05em{\sc i\kern-.025em b}\kern-.08em
    T\kern-.1667em\lower.7ex\hbox{E}\kern-.125emX}}
\begin{document}

\title{Reliability is Blind: \sysname \\ for Decentralized Computing Marketplaces \\ without Individual Behavior Information
\thanks{This work was supported by a French government grant managed by the Agence Nationale de la Recherche under the ANR Labcom program, reference ``ANR-21-LCV1-0012''.}
}

\newboolean{anon}
\setboolean{anon}{false}
\newboolean{oneline}
\setboolean{oneline}{true}
\ifthenelse{\boolean{anon}}{
    \author{\IEEEauthorblockN{Anonymous Authors}
    Submission ID: XX
    }
}
{
    \ifthenelse{\boolean{oneline}}{
        \author{
            \IEEEauthorblockN{Henry Mont\text{*}, Matthieu Bettinger\text{*}\textsuperscript{§}, Sonia Ben Mokhtar\text{*}, Anthony Simonet-Boulogne\textsuperscript{\textdagger}}
            \IEEEauthorblockA{\text{*}INSA Lyon, CNRS, Universite Claude Bernard Lyon 1, LIRIS, UMR5205, 69621 Villeurbanne, France\\
            \{given-name\}.\{surname\}@insa-lyon.fr \textsuperscript{§}Corresponding author}
            \IEEEauthorblockA{\textsuperscript{\textdagger}iExec Blockchain Tech, 69008 Lyon, France
            \{given-name\}.\{surname\}@iex.ec
            }
            Accepted for publication in the \emph{2025 20th European Dependable Computing Conference Companion Proceedings (EDCC-C)}\\
            }
    }
    {
        \author{\IEEEauthorblockN{Henry Mont}
        \IEEEauthorblockA{\textit{LIRIS-DRIM INSA Lyon} \\
        Lyon, France\\
        henry.mont@insa-lyon.fr
        }
        \and
        \IEEEauthorblockN{Matthieu Bettinger}
        \IEEEauthorblockA{\textit{LIRIS-DRIM INSA Lyon} \\
        Lyon, France \\
        matthieu.bettinger@insa-lyon.fr}
        \and
        \IEEEauthorblockN{Sonia Ben Mokhtar}
        \IEEEauthorblockA{\textit{LIRIS-DRIM CNRS} \\
        Lyon, France \\
        sonia.ben-mokhtar@cnrs.fr}
        \and
        \IEEEauthorblockN{Anthony Simonet-Boulogne}
        \IEEEauthorblockA{\textit{iExec Blockchain Tech} \\
        Lyon, France \\
        anthony.simonet-boulogne@iex.ec}
        }
    }
}

\maketitle

\begin{abstract}
    In decentralized cloud computing marketplaces, ensuring fair and efficient interactions among asset providers and end-users is crucial.
    A key concern is meeting agreed-upon service-level objectives like the service's reliability. 
    In this decentralized context, traditional mechanisms often fail to address the complexity of task failures, due to limited available and trustworthy insights into these independent actors' individual behavior. 
    This paper proposes a collective incentive mechanism that blindly punishes all involved parties when a task fails. 
    Based on ruin theory, we show that \sysname improve behavior in the marketplace by creating a disincentive for faults and misbehavior even when the parties at fault are unknown, in turn leading to a more robust marketplace. 
    Simulations for small and large pools of marketplace assets show that \sysname enable to meet or exceed a reliability target, i.e., the success-rate of tasks run using marketplace assets, by eventually discarding failure-prone assets while preserving reliable ones.
\end{abstract}


\section{Introduction}

Centralized cloud service providers shoulder the burden of maintaining and monitoring the quality-of-service of the infrastructure they offer to their tenants, who trust them to be truthful in their service health measurements, to uphold their privacy policies, and to deliver compensation in case of SLA violations.
Decentralized marketplaces, like iExec~\cite{iexecwp}, Ocean Protocol~\cite{oceanwp}, or Secret Network~\cite{secretwp}, can, depending on the platform, propose multiple computing asset types, for example datasets, applications, machine learning models, or servers, for others to match and use.
An important goal for asset providers in those marketplaces is to retain ownership over their assets: they enact usage rules themselves and keep assets confidential (e.g., datasets or models), respectively using blockchain and usually either Secure Multiparty Computations (SMPC) or Trusted Execution Environments (TEEs).
However, this multiplication of independent actors involved in offering a common service (e.g., an application processing a dataset on a server) complexifies the monitoring of executed tasks, as well as enforcing accountability, in particular assigning blame upon failures.
Some failures may have multiple possible root causes stemming from distinct assets, or may be triggered purposely by a malicious actor.
What is more, computing assets that provide the service are heterogeneous in both types and features; they may additionally be updated over time.
For all these reasons, systematic asset auditing or fine-grained monitoring in decentralized systems can reveal complex and costly~\cite{goeschl_non-strategic_2013}.
Therefore, in this paper, we operate in a setup without (trustworthy and available) individual behaviour information: we only assume knowing the task status and who contributed to it (through blockchains), whereas related work usually assumes the existence of some individual contribution information~\cite{hasan_privacy-preserving_2022,kaur_brief_2021, ramachandran_towards_2018,dellarocas2001analyzing}.

\begin{figure}
    \centering
    \includegraphics[trim=2mm 2mm 0 0mm,scale=0.6]{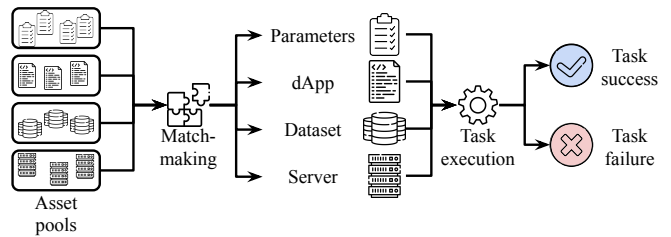}
    \caption{Decentralized cloud computing pipeline example.}
    \label{fig:sysmodel}
    \vspace{-0.5cm}
\end{figure}

\begin{figure*}
    \centering
    \begin{subfigure}{0.32\textwidth}
        \centering
        \includegraphics[scale=1,page=1]{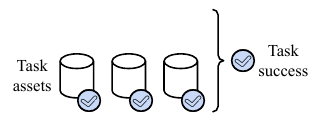}
        \subcaption{Successful task execution}
        \label{fig:fault_free}
    \end{subfigure}
    \begin{subfigure}{0.32\textwidth}
        \centering
        \includegraphics[scale=1,page=2]{\figpathattack}
        \subcaption{Task failure due to an individual failure}
        \label{fig:indiv_fail}
    \end{subfigure}
    \begin{subfigure}{0.32\textwidth}
        \centering
        \includegraphics[scale=1,page=3]{\figpathattack}
        \subcaption{Task failure due to a combination failure}
        \label{fig:group_fail}
    \end{subfigure}
    \vspace{-0cm}
    \caption{Failure model in decentralized cloud computing.}
    \label{fig:attack_model}
    \vspace{-0.25cm}
\end{figure*}

We abstract the decentralized marketplace pipeline as illustrated in \Cref{fig:sysmodel}.
Assets of different types are matched together to execute a given task logic, which can either succeed or fail.
A task fails if at least an individual asset caused a failure (\Cref{fig:indiv_fail}, e.g., due to a bug or to unavailability) or if the combination of a subset of assets did so (\Cref{fig:group_fail}, e.g., a server unable to store a large dataset).
Otherwise, the task succeeds (\Cref{fig:fault_free}).
Failures may occur systematically, all the time or under given conditions, including malicious behaviour, or probabilistically.
In all cases, root causes are considered unknown, so there are no evidently guilty assets.

Game-theoretically~\cite{gao_collective_2015} and empirically~\cite{duell_cooperation_2024}, collective punishment can be more effective than individual punishment in promoting cooperation within groups (of humans or animals).
We apply this insight to decentralized marketplaces of computing assets, where the latter are similarly put together and then interact to perform a given task.
While some assets are a priori passive (e.g., datasets), asset owners can, purposely or not, lead by proxy their assets to deviate from expected behaviour, through bugs, lack of robustness, or malicious intent.

In this paper, we propose a Collective Incentive mechanism where all involved parties in a task are blindly rewarded or punished after their task respectively succeeds or fails.
We formalize using ruin theory~\cite{asmussen_ruin_2010} how to meet a given reliability target, i.e., a task success-rate, then evaluate the protocol's behaviour using simulations in Python (source code available~\cite{DCMIcode}), exploring the impact of system and solution hyperparameters, with both small and large populations of assets that can be matched together for tasks.
Using bounded stakes, \sysname succeed in maintaining or exceeding a target success-rate by ruining failure-prone assets, while sparing reliable assets. 
Additionally, using a reputation score as a selection rate for participation in tasks further improves the task success-rate, as well as precision in preserving reliable assets, by discrediting failure-prone ones.

This paper is structured as follows: \Cref{sec:solution} describes the collective incentive mechanism and how to calibrate it to a chosen task success-rate; \Cref{sec:results,sec:discussion} respectively detail empiric results and discuss future work based on current observations; finally, \Cref{sec:related_work} presents related work and \Cref{sec:conclusion} concludes the paper.

\section{Collective incentives}\label{sec:solution}


Collective incentives operate on the principle that every asset involved in a task faces penalties upon the task's failure.
We detail in this section how to link asset and task failures to the incentive mechanism, in order to drive the system's reliability towards a given target value or under a bound.

\subsection{Quantifying failures}

Let $S_{0}$ represent the initial stake backing an asset $a$, $P$ the penalty imposed per task failure, and $L_{a}$ the expected loss of stake per task.
Let each asset $a$, as well as each combination of assets $A=\{a_{i},\dots,a_{j}\}$, have an intrinsic failure rate, denoted as $F_{a}$ (resp. $F_{A}$).
Failure rates are assumed to follow a long-tailed distribution, where most assets have low failure rates, while a small number of assets are failure-prone, e.g., due to a lack of robustness.
Given a task $t$ and the set of assets $\mathcal{A}_{t}$ involved in executing $t$, task $t$'s failure probability $F_{t}$ depends on the failure rates of elements in the powerset of $\mathcal{A}_{t}$, $\mathbb{P}(\mathcal{A}_{t})$:
\begin{equation}
    F_{t} = 1 - \prod_{A\in \mathbb{P}(\mathcal{A}_{t})}(1 - F_{A})^{|\mathbb{P}(\mathcal{A}_{t})|}
    \label{eq:failure-probability}
\end{equation}
with $F_{A}$ the failure rate of the combination of assets in $A\subseteq\mathcal{A}$ if $|A|\geq 1$ (\Cref{fig:group_fail}), or that of an individual asset if $|A|=1$ (\Cref{fig:indiv_fail}).
Now, given a target task failure-rate $F_{t}^{\mathit{target}}$, a corresponding target asset failure-rate $F_{a}^{\mathit{target}}$ can be determined using \Cref{eq:failure-probability}.
For example, considering the case without combination failures, we have:
\begin{equation}
    F_{a}^{\mathit{target}}=(1-F_{t}^{\mathit{target}})^{1/|\mathcal{A}_{t}|}
\end{equation}

\subsection{Tuning losses}
Knowing how to quantify failure rates, we formalize how much assets expect to lose or gain while executing tasks.
In the general case, the expected loss of stake for an asset $a\in \mathcal{A}_{t}$ per failed task $t$ is:
\begin{equation}
    L_{a} = F_{t} P
    \label{eq:loss-per-task}
\end{equation}


In \Cref{eq:loss-per-task}, as long as there exist penalties $P>0$ or tasks do not have a perfect success rate $F_{t}=0$, $L_{a}>0$ and so $a$ can expect nothing but eventual ruin.
The counterbalance to penalties $P$ is a reward mechanism that provides assets with a partial stake recovery $R$ upon successful task completion.
The expected loss with both penalties and rewards is:
\begin{equation}
    L_{a} = F_{t} P - (1 - F_{t}) R
    \label{eq:net-loss-per-task}
\end{equation}


For an asset to avoid financial ruin, $L_{a}$ should be non-positive, which, using \Cref{eq:net-loss-per-task}, leads to:
\begin{equation}
    F_{t} \leq \frac{R}{P + R}
    \label{eq:loss-inequality-simplified}
\end{equation}

This determines that assets with a task failure rate below $F_{t}^{\mathit{target}}=\frac{R}{P + R}$ can expect to remain profitable, while those above it will eventually face financial ruin.

Finally, to enforce a certain system reliability, we define a desired system failure threshold $F_{t}^{\mathit{target}}$ above which the system is unprofitable for failure-prone assets.
For that purpose, we set the recovery amount $R$ with respect to penalty $P$ such that $F_{t}=F_{t}^{\mathit{target}}$. 
With \Cref{eq:loss-inequality-simplified}, we obtain $R$ and $L_{a}$ as:
\begin{equation}
    R = \frac{F_{t}^{\mathit{target}} P}{1 - F_{t}^{\mathit{target}}}
    \label{eq:recovery-qos}
\end{equation}
\begin{equation}
    L_{a}(F_{t}) = F_{t} \frac{P}{1 - F_{t}^{\mathit{target}}} - \frac{F_{t}^{\mathit{target}} P}{1 - F_{t}^{\mathit{target}}}
    \label{eq:net-loss-qos-expanded}
\end{equation}
$L_{a}$ is positive when $F_{t}>F_{t}^{\mathit{target}}$, tending to deplete $a$'s stake, and non-positive for lower failure rates, replenishing $a$'s stake. 

\section{Evaluation}\label{sec:results}


In this section, we evaluate the extent to which \sysname improve task success rate and survivability of reliable assets in the marketplace.

\subsection{Research questions}

In more detail, we analyze the behaviour of \sysname as well as baselines along the following questions: 

\begin{enumerate}
    \item How much stake is necessary for a reliable asset to survive with a given probability in a system using \sysname?
    \item To what extent can the incentive mechanism keep over a minimum reliability target?
    \item What is the impact of the incentive mechanism on filtering failure-prone assets out of the system? on ensuring survival of reliable assets?
\end{enumerate}

\subsection{Competitors and metrics}


We evaluate \sysname (referred to as ``Coll'') in a context without individual behaviour information, against a baseline system ``Free'' without rewards nor penalties. 

\sysname is further analyzed in three configurations: 
``Coll-Stake'' with stake backing assets and asset removal upon ruin; 
``Coll-Rep'' with remaining asset stake as a reliability or reputation score, which weighs that asset's probability to be selected for tasks, without asset removal upon ruin; 
and finally ``Coll-SR'' which combines both asset removal and stake-weighted asset selection.

We measure the failure rate of tasks executed in the marketplace, as well as the asset removal precision of the mechanisms described above, when applicable.
A true positive is an asset whose stake $S$ was depleted (i.e., $S<P$) and had an asset failure rate $F_{a}$ above the per-asset threshold, i.e., $F_{a}>F^{\mathit{target}}_{a}$.
Conversely, true negatives are assets whose stake remains greater or equal to $P$ during the duration of an experiment and who satisfy $F_{a}\leq F^{\mathit{target}}_{a}$.
False positives and negatives are assets who satisfy the corresponding definition above about their stake but not their asset failure rate.

\subsection{Simulation configuration}

We simulate in Python a decentralized computing marketplace where $N_{t}$ task executions take place~\cite{DCMIcode}.
In a real-world scenario, e.g., on the iExec platform~\cite{iexecwp}, \sysname would be implemented on Smart Contracts (for transparency and immutability).
For simplicity in presented experiments, we consider that there are always 4 assets involved in any given task (i.e., like on iExec).
There are $N_{a}$ assets available initially, some of which may be removed from the pool over time by an incentive mechanism upon ruin.
We allow new assets to be added to the marketplace to replace removed assets, every $T=500$ tasks in presented experiments.
Each asset $a$ has an individual failure rate $F_{a}$ drawn from a power-law distribution of parameter $\alpha$.
In this version of the paper, we only consider individual asset failures, not asset combination failures, i.e., in \Cref{eq:failure-probability}, $F_{A}=0$ when $|A|>1$.

Finally, we have three hyper-parameters for \sysname: the initial per-asset stake $S_{0}$, the penalty $P$, and the target system reliability $F_{t}^{\mathit{target}}$.
We consider in the following that $S_{0}=xP$, with $x\geq0$ and we use a unit penalty $P=1$.
With a fixed $P$, reward $R$ is a function of $F_{t}^{\mathit{target}}$, and loss $L_{a}$ additionally of $F_{t}$.
Both are directly proportional to $P$, so results with $P\neq1$ can be deduced easily.

\subsection{Results}

We now present empiric results to our research questions.

\subsubsection{Initial stake $S_{0}$ and reliable asset survivability}

\begin{figure}
    \centering
    \includegraphics[trim=2mm 2mm 2mm 2mm,scale=0.75]{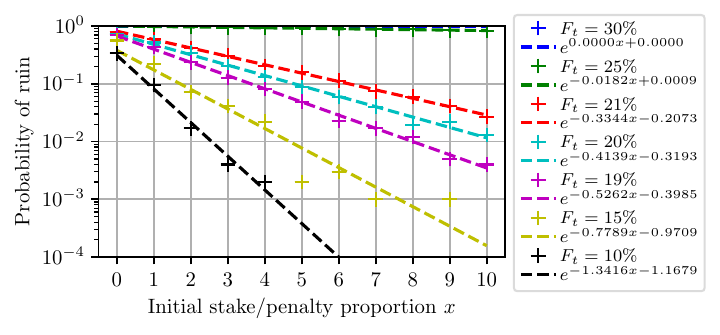}
    \caption{Asset ruin probability depending on initial stake $S_{0}=xP$ and experienced task failure rate $F_{t}$, given $F_{t}^{\mathit{target}}=20\%$.}
    \label{fig:ruin-probabilities}
    \vspace{-0.4cm}
\end{figure}

First, we investigate the impact of the initial stake $S_{0}$ on an asset's probability of ruin.
Indeed, even if an asset's individual failure rate $F_{a}$ is below the target $F_{t}^{\mathit{target}}$, there remains the probability of suffering sufficient failures over a bounded timeframe such that the stake becomes depleted.
The Cramér-Lundberg model~\cite{asmussen_ruin_2010} captures these fluctuations and enables to estimate the probability of ruin over a horizon, in our case of $N_{t}$ executed tasks.
\Cref{fig:ruin-probabilities} shows the ruin probability for different values of the stake/penalty ratio $x$ and of $F_{t}$, with $F_{t}^{\mathit{target}}=20\%$ and over a horizon of $N_{t}=10^{4}$ tasks.
Given $F_{t}$, the probability of ruin decreases exponentially with $x$.
The smaller $F_{t}$ compared to $F_{t}^{\mathit{target}}$, the steeper the decrease, while increasingly higher $F_{t}$ than $F_{t}^{\mathit{target}}$ tend towards a constant probability of ruin of $1$.
As a consequence, the better the asset's reliability, the less stake is required to expect its survival in the marketplace with a given probability.
In the following, we set $S_{0}=10P$ for all assets.

\subsubsection{Task reliability service-level objectives}

\begin{figure}
    \centering
    \includegraphics[trim=2mm 4mm 2mm 1mm,scale=0.8]{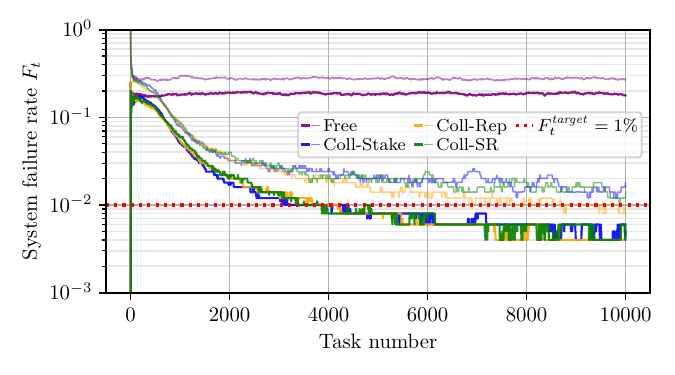}
    \caption{Median (dark) and $95^{\mathit{th}}$ percentile (light) task failure rates over 100 runs for different incentive mechanisms (using per-run task failure-rates computed as moving averages with a window size of 100 tasks).
    $N_{t}=10^{4}$ tasks each involving 4 assets are executed per run, with $N_{a}=100$, $\alpha=0.05$ ($\mathit{mean}(F_{a})=4.8\%$), $S_{0}=10$, and $F_{t}^{target}=1\%$.
    }
    \label{fig:marketplace-failure-rate}
    \vspace{-0.4cm}
\end{figure}

We now evaluate the extent to which collective incentive mechanisms can keep over a minimum reliability bound (or, equivalently, under a maximum failure rate) in \Cref{fig:marketplace-failure-rate}.
Without incentives, in the ``Free'' system, the task failure rate remains between around 10\% and 30\%.
With \sysname, after around 4000 tasks, around 95\% of the runs have a task failure rate below 2\%, the target used by the mechanisms being $F_{t}^{\mathit{target}}=1\%$.
Therefore, incentive mechanisms can effectively maintain a low task failure rate in the marketplace.
In the next experiments, we will observe how changing the hyperparameters of the marketplace or of the incentive mechanism affects the task failure rate.

\subsubsection{Asset discrimination on reliability}

\begin{figure*}[h]
    \centering
    \begin{subfigure}{0.49\textwidth}
        \centering
        \includegraphics[trim=8mm 4mm 6mm 4mm,scale=0.8]{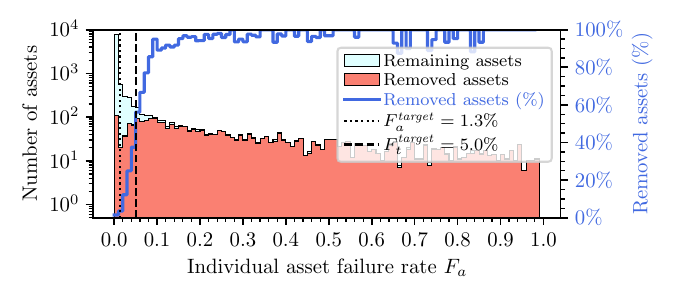}
        \caption{Asset survival with respect to individual failure rate $F_{a}$.}
        \label{fig:individual-failure-rate}
    \end{subfigure}
    \begin{subfigure}{0.49\textwidth}
        \centering
        \includegraphics[trim=8mm 4mm 6mm 4mm,scale=0.8]{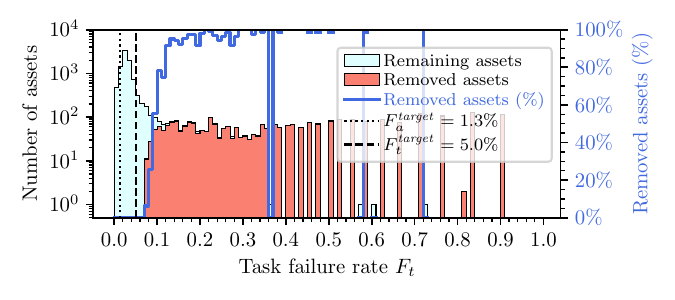}
        \caption{Asset survival with respect to experienced task failure rate $F_{t}$.}
        \label{fig:effective-failure-rate}
    \end{subfigure}
    \caption{Stacked bars of removed and remaining assets using ``Coll-Stake'' after $N_{t}=10^{4}$ tasks each involving 4 assets, averaged over 100 simulation runs, with $N_{a}=100$, $\alpha=0.1$ ($\mathit{mean}(F_{a})=9.2\%$), $S_{0}=10$, $F_{t}^{target}=5\%$.}
    \label{fig:failure-rate}
\end{figure*}

A concern with staking-based mechanisms is to unjustly ruin and remove reliable assets from the marketplace.
In \Cref{fig:failure-rate}, after using ``Coll-SR'' on marketplaces for $N_{t}=10^{4}$ tasks, 
we observe the populations of removed (lower, red) and remaining (upper, light-blue) assets and analyze their individual and task failure rates compared to the respective targets $F_{a}^{\mathit{target}}$ and $F_{t}^{\mathit{target}}$.
The blue curves in \Cref{fig:individual-failure-rate,fig:effective-failure-rate} measure the proportion of removed assets among the total number of assets in the same failure-rate bin.
In \Cref{fig:individual-failure-rate}, this removal proportion increases approximately linearly after $F_{a}^{\mathit{target}}$, then flattens and tends to remain around 100\% for higher $F_{a}$.
The progressive increase can be explained by the fact that an asset with $F_{a}^{\mathit{target}}<F_{a}<F_{t}^{\mathit{target}}$ can still experience an average task failure rate lower than $F_{t}^{\mathit{target}}$ if it is matched with really reliable assets.
We indeed see in \Cref{fig:effective-failure-rate}, which illustrates task failure rates experienced by assets, that the asset removal proportion only increases after the $F_{t}^{\mathit{target}}$ threshold.
Therefore, \sysname can effectively filter out failure-prone assets.

\begin{table*}[h]
    \resizebox{\textwidth}{!}{\begin{tabular}{r|cc|cc|cc|cc|cc|cc|cc|cc|cc}
        $N_{t}$       & \multicolumn{2}{c|}{10000}                                   & \multicolumn{2}{c|}{10000}                                   & \multicolumn{2}{c|}{10000}                                   & \multicolumn{2}{c|}{10000}                                  & \multicolumn{2}{c|}{10000}                                   & \multicolumn{2}{c|}{\cellcolor[HTML]{DAE8FC}10000}                                   & \multicolumn{2}{c|}{10000}                                   & \multicolumn{2}{c|}{10000}                                  & \multicolumn{2}{c}{\cellcolor[HTML]{DAE8FC}30000}           \\
        $N_{a}$       & \multicolumn{2}{c|}{\cellcolor[HTML]{DAE8FC}10}              & \multicolumn{2}{c|}{100}                                     & \multicolumn{2}{c|}{100}                                     & \multicolumn{2}{c|}{100}                                    & \multicolumn{2}{c|}{100}                                     & \multicolumn{2}{c|}{\cellcolor[HTML]{DAE8FC}100}                                     & \multicolumn{2}{c|}{100}                                     & \multicolumn{2}{c|}{\cellcolor[HTML]{DAE8FC}1000}           & \multicolumn{2}{c}{\cellcolor[HTML]{DAE8FC}1000}            \\
        $S_{0}$       & \multicolumn{2}{c|}{10}                                      & \multicolumn{2}{c|}{\cellcolor[HTML]{DAE8FC}2}               & \multicolumn{2}{c|}{\cellcolor[HTML]{DAE8FC}5}               & \multicolumn{2}{c|}{10}                                     & \multicolumn{2}{c|}{10}                                      & \multicolumn{2}{c|}{\cellcolor[HTML]{DAE8FC}10}                                      & \multicolumn{2}{c|}{10}                                      & \multicolumn{2}{c|}{10}                                     & \multicolumn{2}{c}{10}                                      \\
        $\alpha$      & \multicolumn{2}{c|}{0.05}                                    & \multicolumn{2}{c|}{0.05}                                    & \multicolumn{2}{c|}{0.05}                                    & \multicolumn{2}{c|}{\cellcolor[HTML]{DAE8FC}0.01}           & \multicolumn{2}{c|}{\cellcolor[HTML]{DAE8FC}0.1}             & \multicolumn{2}{c|}{\cellcolor[HTML]{DAE8FC}0.05}                                    & \multicolumn{2}{c|}{0.05}                                    & \multicolumn{2}{c|}{0.05}                                   & \multicolumn{2}{c}{0.05}                                    \\
        $F_{t}^{\mathit{target}}$       & \multicolumn{2}{c|}{0.01}                  & \multicolumn{2}{c|}{0.01}                                    & \multicolumn{2}{c|}{0.01}                                    & \multicolumn{2}{c|}{0.01}                                   & \multicolumn{2}{c|}{0.01}                                    & \multicolumn{2}{c|}{\cellcolor[HTML]{DAE8FC}0.01}                                    & \multicolumn{2}{c|}{\cellcolor[HTML]{DAE8FC}0.05}                                    & \multicolumn{2}{c|}{0.01}                                   & \multicolumn{2}{c}{0.01}                                    \\
        \hline
        Coll-Stake & 91.0                         & 0.59                         & \cellcolor[HTML]{DAE8FC}52.3 & 0.72                         & 76.9                         & 1.50                         & \cellcolor[HTML]{DAE8FC}100 & 0.56                         & 74.7 & \cellcolor[HTML]{DAE8FC}3.75                                 & 92.6               & 2.46                         & 99.4                         & 2.85                               & 99.0                   & 10.2 & 95.3                         & 4.72 \\
        Coll-SR    & \cellcolor[HTML]{DAE8FC}95.2 & \cellcolor[HTML]{DAE8FC}0.54 & 31.3 & \cellcolor[HTML]{DAE8FC}0.49 & \cellcolor[HTML]{DAE8FC}85.5 & \cellcolor[HTML]{DAE8FC}1.40                     & \cellcolor[HTML]{DAE8FC}100 & \cellcolor[HTML]{DAE8FC}0.51 & \cellcolor[HTML]{DAE8FC}78.5 & 3.84                         & \cellcolor[HTML]{DAE8FC}98.6 & \cellcolor[HTML]{DAE8FC}2.37   & \cellcolor[HTML]{DAE8FC}99.7 & \cellcolor[HTML]{DAE8FC}2.81   & \cellcolor[HTML]{DAE8FC}100 & \cellcolor[HTML]{DAE8FC}9.48 & \cellcolor[HTML]{DAE8FC}99.7 & \cellcolor[HTML]{DAE8FC}4.59
    \end{tabular}
    }
    \caption{Percentages of asset removal precision (left) and task failure rate $F_{t}$ (right), for different market and incentive hyperparameter valuations.
    For hyperparameter rows (first five rows), the fully colored column is the reference and the other colored cells are changes from reference.
    For incentive mechanism rows, blue cells highlight the best scores per configuration.}
    \label{tab:precision-failure-rate}
    \vspace{-0.25cm}
\end{table*}

Finally, to assess whether \sysname also both preserve reliable assets and high reliability simultaneously, in \Cref{tab:precision-failure-rate}, we measure the precision of asset removal and average task failure rate, for different hyperparameter valuations.
We compare the results of ``Coll-Stake'' and ``Coll-SR'', which both use staking and asset removal, while ``Coll-SR'' additionally preferentially selects assets with high stakes.
Mechanisms without asset removal (``Coll-Rep'' and ``Free'') do not have a removal precision value, so they are not included \Cref{tab:precision-failure-rate}.
Varying hyperparameters, we observe that ``Coll-SR'' outperforms ``Coll-Stake'' in most cases both in terms of asset removal precision and task failure rate.
In all cases except for the second to last configuration, the task failure rate is under 1\% to 5\%.
The reason for this exception is that $N_{a}$ is significantly higher in the last two configurations: the space to randomly select assets is larger, so the system did not converge at $N_{t}=10^{4}$, but reached below 5\% failure-rate after $N_{t}=3\cdot10^{4}$ tasks.
The second and third columns, where $S_{0}$ varies, corroborate the results from \Cref{fig:ruin-probabilities}: the higher the initial stake, the fewer reliable assets are removed from the marketplace.
Regarding the performance with respect to $\alpha$, which adjusts how failure-prone assets are, and $F_{t}^{\mathit{target}}$, we observe than even with a high $\alpha=0.1$ and low $F_{t}^{\mathit{target}}=1\%$ ($5^{\mathit{th}}$ column), the system can still maintain a low task failure rate.
Note that $\alpha$ determines $F_{a}$, not directly $F_{t}$: we have for $\alpha=0.1$ the corresponding $\mathit{mean}(F_{a})=9.2\%$ and using \Cref{eq:failure-probability}, $F_{t}$ is around $32\%$, significantly more than $F_{t}=3.8\%$ in \Cref{tab:precision-failure-rate}'s $5^{\mathit{th}}$ column.

\section{Discussion}\label{sec:discussion}

In this section, we discuss the implications of our results, then suggest improvements and extensions to our work.

\subsection{Results analysis}

We have shown that \sysname, i.e., blindly punishing and rewarding all actors involved in a task for its failure or success, can improve the reliability of decentralized cloud computing marketplaces, in non-adversarial environments with individual failures.
\sysname can both reduce the task failure rate while avoiding ruin for reliable actors.
Overall, results suggest that ``Coll-SR'', combining both reputation and staking with \sysname, improves both overall task reliability and precision in preserving already reliable assets.




\subsection{Extensions upon current work}

As mentioned in our experimental protocol, we have not yet explored the incentive's performance with failures stemming from asset combinations.
We expect that an additional mechanism next to the collective incentive system will be necessary to handle these cases.
Indeed, in that case, instead of removing assets from the marketplace, these assets should avoid being involved in combinations that were prone to fail in the past.
Aral et al.~\cite{aral_reliability_2020} propose a selection mechanism using Bayesian networks, which predicts combination reliability based on past performance, but their approach is an NP-Complete task.
In between their work and ours, researching an efficient approach that approximates the problem is promising.

Additionally, we have yet to investigate the system's resilience in adversarial environments.
Attacks on Web3, staking and reputation systems, are well-explored~\cite{hasan_privacy-preserving_2022}: future work includes applying these attacks to our system, evaluating its robustness, and proposing countermeasures where necessary.
\section{Related work}\label{sec:related_work}
The issue addressed by our \sysname spans several research domains, which we present and compare to our approach in this section.

\subsection{Decentralized cloud computing}
Decentralized cloud computing leverages blockchain technology to let any owner of monetizable computing assets provide them to other, while keeping control on how these assets are used.
However, monitoring reliability and security amongst actors, assets, and executions, is challenging in a decentralized setting.
For instance, the iExec marketplace~\cite{iexecwp} facilitates the decentralized buying and selling of applications, datasets, and computing power, using smart contracts to manage interactions without the need for a central authority.
While it uses Trusted Execution Environments (TEEs), which can provide trusted execution of code, including monitoring, it is not easily actionable for monitoring arbitrary and confidential code.
Moreover, TEEs are surrounded by a potentially malicious host, which can manipulate procotol interactions outside of the TEE's monitoring scope.

\subsection{Incentive mechanisms}
Incentive mechanisms are essential for promoting cooperation and active participation in decentralized systems.
Huang et al.~\cite{huang_survey_2019} provide a comprehensive survey on blockchain incentive mechanisms, focusing on the architecture and goals of blockchain's incentive layer.
They discuss the issuance of tokens concerning computation, storage, and transmission, analyzing the rationality of token allocation paths.
While their work underscores the importance of incentives in sustaining blockchain networks, it does not address the challenges of systems with incomplete contribution information.

Maddikunta et al.~\cite{maddikunta_incentive_2022} review incentive techniques in Internet of Things (IoT), highlighting the roles of blockchain, game theory, and artificial intelligence (AI) in stimulating device participation.
Covering multiple applications, their focus remains on individual device incentives rather than collective mechanisms, and they do not cover the distribution of penalties among a group to enhance system reliability.

\subsection{Monitorless incentivization}
In scenarios with human or animal agents where individual contributions are not easily discernible, approaches rely on the overall transaction outcome rather than individual actions to impose sanctions or rewards.
Socio-economic studies highlight the problem of costly monitoring~\cite{goeschl_non-strategic_2013} and investigate the effectiveness of collective punishment in promoting cooperation in settings with imperfect monitoring~\cite{duell_cooperation_2024,gao_collective_2015,fatas_blind_2010}.
Goeschl and Jarke~\cite{goeschl_non-strategic_2013} study environments with costly monitoring, showing that players often resort to non-strategic or blind punishment when individual actions are not observable.
Although not directly applicable to our specific decentralized system, these studies reflect viable strategies for incentivizing cooperation and accountability in environments with limited information.

\subsection{Token-Curated Registries}
Token Curated Registries (TCR), can be used as a way to maintain lists of reliable and trustworthy buyers, sellers or products.
Ramachandran et al.~\cite{ramachandran_towards_2018} discuss the use of TCRs to curate reliable data sources in smart city data marketplaces, emphasizing their potential to enhance data quality and trustworthiness.
This approach involves data providers staking tokens to apply for inclusion in the registry, where existing token holders vote to accept or reject applications.
The economic incentive for token holders is to maintain a high-quality list, thereby increasing the value of the tokens they hold.
TCRs provide transparency and trust, enabled through decentralization using blockchain~\cite{kaur_brief_2021}.
However, TCRs primarily address single-agent transactions involving one buyer and one seller, limiting their applicability in more complex, multi-agent scenarios.
Additionally, the necessity for participants to manage registries and provide incentives adds complexity and overhead.

\subsection{Reputation mechanisms}
Reputation mechanisms are crucial for building trust and cooperation in online communities by collecting and publishing user feedback, aiming for more informed future decisions~\cite{dellarocas2006reputation}.
In a decentralized cloud computing marketplace, these mechanisms enable participants to select reliable and trustworthy partners to achieve their goals.
Sellers in marketplaces like eBay can advertise any quality, but their true quality is eventually revealed through buyer ratings~\cite{dellarocas2001analyzing}.
However, incomplete information can hinder buyers from accurately assessing seller quality, leading to inefficiencies, especially in decentralized and multi-agent transactions where the buyer may not know who is responsible for unsatisfactory services.
Hasan et al.'s survey highlights privacy-preserving reputation systems and their key characteristics~\cite{hasan_privacy-preserving_2022}.
These systems are unsuitable for our context as they either rely on a trusted third party or assume the buyer can provide an informed review of the transaction, which is not feasible in our situation.

\subsection{Blended systems}
Blended mechanisms such as the one proposed by Li et al.~\cite{li_game-theoretic_2012} offer promising strategies for fostering cooperation and trust in various networks.
These mechanisms leverage the strengths of both reputation-based and price-based systems to create a more robust incentive structure, as we do with \sysname.
For instance, such a system might adjust the cost of services based on the reputation of the nodes involved.
This dual approach reinforces cooperative behavior by tying these rewards to reputation, making it harder for selfish nodes to exploit the system.
However, while blended mechanisms provide a stronger framework for incentivizing cooperation, these systems rely on individual contribution information, unavailable in our context.

\subsection{Bayesian network diagnostic / predictive reasoning}
In our Web3 computing marketplace context, where resource providers are distributed, geographically diverse, and often uncoordinated, Bayesian networks (BNs) offer a robust framework for predicting the reliability of providers by modeling dependencies between them.
Aral et al.~\cite{aral_reliability_2020} explore the use of BNs for reliability management in blockchain-based decentralized multi-cloud.
Their solution uses BNs to extract dependencies from historical log traces and identify potential correlations between providers, which can arise due to shared ownership or other underlying factors.
They finally select the most reliable providers based on these correlations.
While effective, applying Bayesian networks in a fully decentralized environment is challenging: Bayesian network inference is known to be NP-complete. 
Hence it is computationally expensive, leading to scalability issues as the size and complexity of the decentralized network grow.

\section{Conclusion}\label{sec:conclusion}

Web3 cloud computing marketplaces open up opportunities to monetize assets while keeping control over them.
However, confidentiality constraints and the multiplication of independent actors in these decentralized environments make it challenging to obtain insights on the health of the system and to find root causes of failures.
In this work, we tackle limited behaviour information about the involved actors by proposing a collective incentive mechanism that punishes and rewards all actors based on their task's success or failure.
We show, in a non-adversarial context with individual failures, that this mechanism can enforce a reliability threshold in decentralized cloud computing marketplaces, using a tuned combination of staking and reputation mechanisms, without ruining reliable actors.
Notably, in presented experiments, our \sysname succeeded in decreasing the task failure rate 5- to 10-fold, while preserving a large majority of already reliable assets.
This approach simplifies the management of task failures and provides a promising solution for Web3 environments where trustworthy information is inherently limited.
\bibliographystyle{ieeetran}
\bibliography{imports/bibliography}

\end{document}